\documentclass[aps,prl,twocolumn,superscriptaddress]{revtex4-2}

\usepackage[utf8]{inputenc}
\usepackage[T1]{fontenc}
\usepackage{amsmath,amssymb,amsfonts}
\usepackage{algorithmic}
\usepackage{hyperref}
\usepackage{epstopdf}
\usepackage{textcomp}
\usepackage{graphicx}
\usepackage{mathrsfs}
\usepackage[vlined,ruled]{algorithm2e}
\usepackage{subfigure}
\usepackage{url}
\usepackage{color}
\usepackage{dsfont}
\usepackage{bbm}
\usepackage{booktabs}
\usepackage{array}
\usepackage[table]{xcolor}

\newcommand{\alias}{C:/Users/yishi/Dropbox/bib/alias}
\newcommand{\New}{C:/Users/yishi/Dropbox/bib/New}
\newcommand{\Main}{C:/Users/yishi/Dropbox/bib/Main}
\newcommand{\FP}{C:/Users/yishi/Dropbox/bib/FP}

\newcommand{\s}{{\rm s}}
\newcommand{\f}{{\rm f}}

\newcommand{\I}{\mathsf{i}}

\newcommand{\RI}{{\rm  I}}

\begin{document}

\title[Phase-Amplitude Coupling]{Phase-Amplitude Coupling in Neuronal Oscillator Networks}

\author{Yuzhen Qin}
\email[]{yuzhenq@ucr.edu}
\affiliation{Department of Mechanical Engineering, University of California, Riverside}

\author{Tommaso Menara}
\affiliation{Department of Mechanical Engineering, University of California, Riverside}

\author{Danielle S. Bassett}
\affiliation{
Department of Physics and Astronomy,
Department of Bioengineering,
Department of Electrical \& Systems Engineering,
Department of Neurology,
Department of Psychiatry, University of Pennsylvania \& the Santa Fe Institute}

\author{Fabio Pasqualetti}
\email[]{fabiopas@engr.ucr.edu}
\affiliation{Department of Mechanical Engineering, University of California, Riverside}



\begin{abstract}
Phase-amplitude coupling (PAC) describes the phenomenon where the power of a high-frequency oscillation evolves with the phase of a low-frequency one. We propose a model that explains the emergence of PAC in two commonly-accepted architectures in the brain, namely, a high-frequency neural oscillation driven by an external low-frequency input and two interacting local oscillations with distinct, locally-generated frequencies. We further propose an interconnection structure for brain regions and demonstrate that low-frequency phase synchrony can integrate high-frequency activities regulated by local PAC and control the direction of information flow across distant regions.
\end{abstract}
\maketitle


Oscillatory activity at multiple frequency bands is widely observed in the brain, among other natural, biological and technological systems. It is thought that synchronization of brain oscillations within the same frequency band supports effective neural communication \cite{PS-PJM:2012}, and extensive studies have investigated this phenomenon \cite{TPA-FTJ:03,NV-VM:13,QY-CM-ABDO-BD-QF:2020,TM-GB-DSB-FP:18}. Yet, there is an emerging body of evidence revealing that brain oscillations in distinct frequency bands are not independent, and they are instead found to interact through cross-frequency coupling~\cite{CRT-EE-DSS:2006,JO-CLL:2007}.

\begin{figure}[t]
	\includegraphics[scale=0.75]{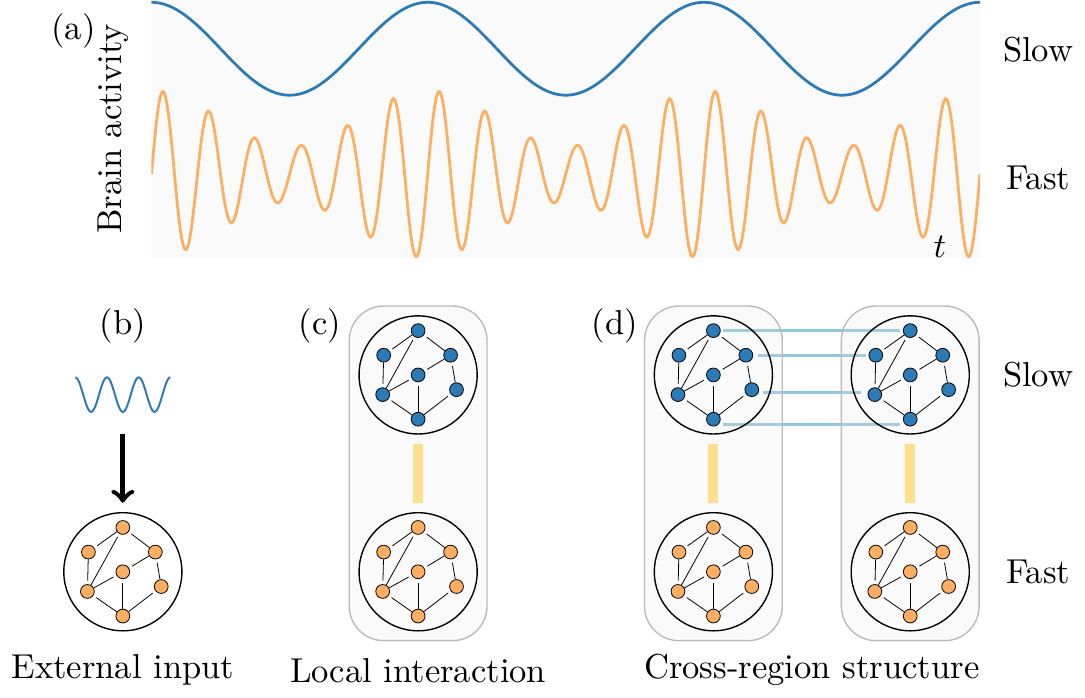}
	\caption{An overview of PAC. (a) Illustration of PAC. (b)-(c) Different architectures for the emergence of PAC: (b) a fast population with an external slow input (e.g., sensori stimulus); (c) two interacting populations with distinct local oscillations (which also constitute a simplified brain region). (d) Cross-region interconnection structure.}
	\label{PAC}
\end{figure}

As one of the most prevalent representations of cross-frequency coupling \cite{CRT-KRT:2010}, phase-amplitude coupling (PAC) occurs when the power (or amplitude) of a high-frequency rhythm is locked to (and often modulated by) the phase of a low-frequency rhythm (see FIG.~\ref{PAC}(a)). Brain signals recorded by various techniques, e.g., local field potential (LFP), electroencephalograph (EEG), and magnetoencephalograph (MEG), have revealed that PAC emerges in numerous brain regions, including auditory and prefrontal cortices \cite{LP-SAS-KKH:2005,VB-VT-ES-WT:2015}, nucleus \cite{CMX-AN:2009}, and hippocampus \cite{AN-HMM-JO:2010}, and plays a crucial role in motor functions \cite{YT-YO:2012} and cognitive processes such as working memory \cite{RF-UP:2014}, attention \cite{LP-KG-MAD:2008}, and learning \cite{TABL-KRW:2009}. 

Despite extensive empirical work on PAC, its underlying mechanisms remain largely unknown. Modeling PAC can help us uncover the conditions under which this phenomenon emerges, shedding light on the working principles of numerous cognitive functions. Most of the existing models focus on the microscopic scale \cite{WJA-BMI:2000,TA-RHG-DT:2007,WP-PAA-BM:2009,FL-KM-HA-GB:2013}, but they fail to capture the behavior of larger neural populations. Some attempts have been made to describe PAC at the population level with the aid of neural mass models \cite{AT-OR-ML:2012,OACE-JMW-BR:2014,JM-PAJ-GOJ:2015,CM-VLC:2017}. Two architectures have been accepted to generate PAC \cite{HA-GAL-FL-GB:2015}: 1) a neural population oscillating at high-frequency (fast) is modulated by an external low-frequency (slow) input (see FIG.~\ref{PAC}(b)); 2) two populations that respectively generate fast and slow oscillations interact with each other (see FIG. \ref{PAC}(c)). Yet, a model capable of assimilating the aforementioned instances with large-scale synchronization of multiple brain regions is still critically missing.

PAC usually does not emerge in the brain as an isolated phenomenon. In fact, regions exhibiting local PAC may also be phase-coupled in low frequency bands. The combination of cross-region phase synchronization and local cross-frequency PAC can thus effectively integrate local information in distant brain regions, which is processed and regulated by high-frequency oscillations \cite{CRT-KRT:2010}. For instance, correlated $\gamma$  rhythms (>30 Hz) between the cortex and striatum are typically established by cross-regional phase synchronization in the $\theta$ band (4-8 Hz) concurrently with $\theta$-$\gamma$ PAC \cite{VNC-EG-SA:2014}. 
Finally, while synchronization phenomena enable effective communication, the question of which regions send and which ones receive such information often remains unanswered. Determining the directionality of information flow is of paramount importance for revealing the spatio-temporal hierarchy of brain processes. For instance, it has been shown that during working memory maintenance the prefrontal cortex receives information from the sensory cortices \cite{LAH-WJD:2015} and sends information to the inferior temporal cortex \cite{DJ-GT-EAK-FU:2017}. 


In this Letter, we propose a neurobiologically plausible model based on the Stuart-Landau (SL) equation that allows for the emergence of  cross-frequency coupling in both architectures depicted in FIG. \ref{PAC}(b)-(c). Further, inspired by previous studies showing that low-frequency oscillations are better suited for establishing long-distance interactions than high-frequency ones \cite{VSA-SJ:2000,VNC-EG-SA:2014}, we provide an interconnection structure (see FIG.~\ref{PAC}(d)) in which cross-region interconnections exist only between slow populations. This structure creates a framework that integrates cross-frequency coupling with same-frequency phase synchronization, and constitutes a building block that can be easily scaled to form large networks. Under this structure, we demonstrate the important role that long-distance same-frequency phase synchrony, together with regional PAC, plays in the coordination of high-frequency local activity and in information routing.

We start by introducing the SL equation:
\begin{align}\label{SL:main}
\dot z=(\sigma+\I  \omega-|z|^2)z,
\end{align}
where $z$ is a complex state ($z=x+\I y$), with $\I=\sqrt{-1}$, and $\sigma$ and $\omega$ are real. The SL oscillator is also known as a Hopf oscillator, because, for $\sigma<0$, $z=0$ is globally stable, and for $\sigma>0$ the model undergoes a Hopf bifurcation with $z$ converging to a circular trajectory (limit cycle) of radius $\sqrt{\sigma}$ and angular frequency $\omega$. The SL oscillator can equivalently be described in polar coordinates by letting $z=re^{\I\theta}$, where $r$ and $\theta$ are referred to as the amplitude and phase angle of the oscillator, respectively. While phase-only models such as the Kuramoto model have been widely used to study synchrony in brain oscillations \cite{ME-PD:18,QY-KY-PO-CM:19,TM-GB-DSB-FP:19b,QY-CM-ABDO-BD-QF:2020,TM-GB-DSB-FP:18}, the SL model, like many other phase-amplitude models \cite{SA-DRM:2020,MMH-EJ:2019}, can capture richer behaviors than phase-only ones.

We use the SL equation to describe the collective dynamics of a single neural population. The SL equation can be derived from the well-established Wilson-Cowan model of interaction between excitatory and inhibitory neural populations \cite{HGS-PW:90}, and from the reduction of a network of inhibitory integrate-and-fire neurons \cite{BN-HV:1999}. Despite its apparent simplicity, the SL equation is a very powerful approximation of neural dynamics around the Hopf bifurcation. In fact, local synaptic coupling can synchronize the firing activity of neurons \cite{HA-GAL-FL-GB:2015}, and then bring the population from a stationary to an oscillatory regime by shifting the parameter $\sigma$ in Eq.~\eqref{SL:main} from negative to positive. The oscillation frequency mainly depends on the intrinsic decay time of inhibition in the population \cite{BC-KN:2005}.

\noindent \emph{Emergence of PAC by exogenous inputs} -- We first show how PAC can arise in a neural population oscillating at a high frequency subject to a slow exogenous input (i.e., the architecture depicted in FIG.~\ref{PAC}(b)). Consider the dynamics of the neural population with an input $u$
\begin{align}\label{SL:main:input}
\dot z_\f=(\delta_\f+\I  \omega_\f-|z_\f|^2)z_\f+z_\f u,
\end{align}
where $z_\f=x_\f+\I y_\f$, and $\delta_\f,v_\f>0$. The input $u$ can represent various sensory signals from sensory cortices \cite{MA-PS-LMK-BN:2008,HA-GAL-FL-GB:2015}.  We let $u$ be a sinusoidal signal of strength $k_\RI$, i.e., $u=k_\RI \sin (\omega_\RI t)$. Consequently, 
\begin{align*}
\dot z_\f=(\delta_\f+k_\RI \sin (\omega_\RI t)+\I  \omega_\f-|z_\f|^2)z_\f.
\end{align*}
Then, $\sigma$ in Eq. \eqref{SL:main} becomes time-dependent, i.e., $\sigma(t)=\delta_\f+k_\RI \sin (\omega_\RI t)$, and the fast amplitude $r_\f$ tends to converge to $\sqrt{\sigma(t)}$, yielding PAC.

\begin{figure}[t]
	\includegraphics[scale=0.7]{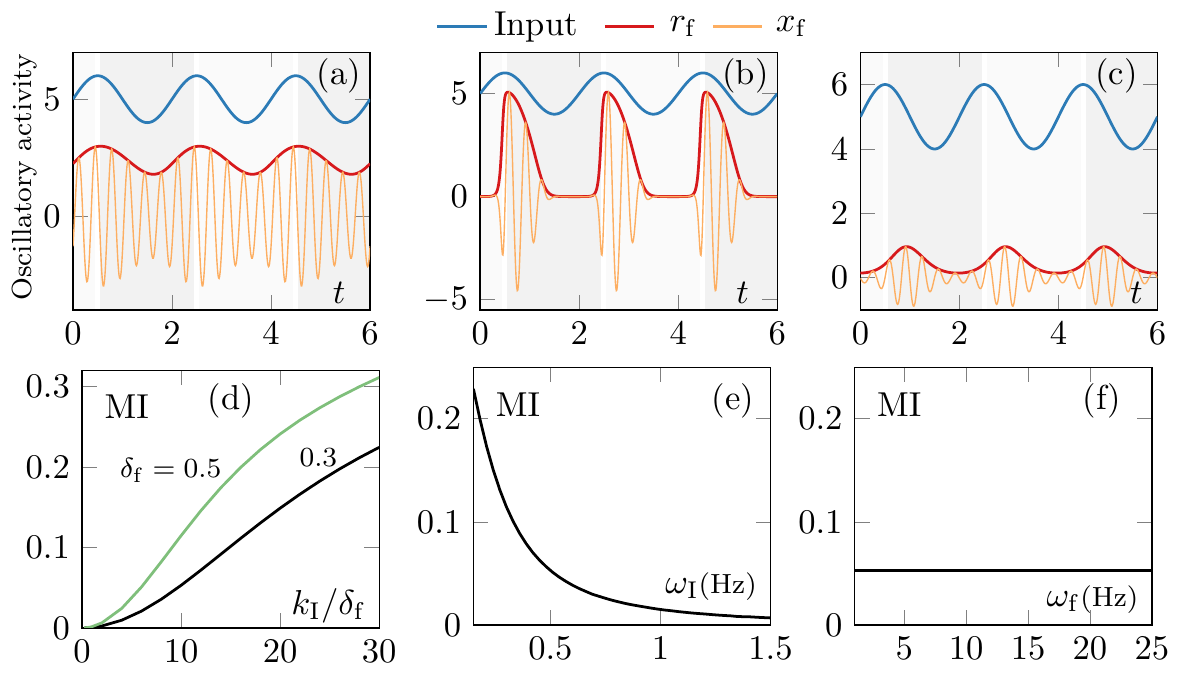}
	\caption{PAC by exogenous inputs. (a)-(b) A small $k_\RI$ yields continuous PAC, while a large one yields intermittent PAC ($\delta_\f=6$, $k_\RI$ is 3 for panel (a) and 18 for panel (b)). (c) $\delta_\f$ in panel (a) is reduced to $0.3$. (d)-(f) The dependence of PAC intensity (measured by the modulation index (MI)) on the input strength (d), the input frequency (e), and the fast frequency (f). (If not specified otherwise, $\delta_\f=0.3$, $k_\RI=3$ $\omega_f=3$ Hz, and $\omega_\RI=0.5$ Hz.)}
	\label{Input}
\end{figure}

Notice that, depending on how the fast oscillation behaves throughout
the slow cycle, PAC can be categorized into continuous and
intermittent. In the former, the fast oscillation is constantly
active, while in the latter, the fast oscillation only appears in a
certain phase interval of the slow cycle \cite{HA-GAL-FL-GB:2015}. The
input magnitude determines which type of PAC will occur. For a weak
input, where $k_\RI<\delta_\f$, $\sigma(t)$ is always positive and the
fast population remains in the oscillatory regime, which implies that
continuous PAC arises (see FIG.~\ref{Input}(a)). Conversely, for a
strong input, where $k_\RI>\delta_\f$, $\sigma(t)$ can be negative in
some interval of the slow cycle. In these intervals, the fast
population is driven out of the oscillatory regime. Thus, $r_\f$ tends
towards $0$, and the fast oscillation may disappear. We
illustrate intermittent PAC for large $k_\RI$ in FIG.~\ref{Input}(b).

One can observe from FIG. \ref{Input}(a)-(b) that the peaks and valleys of the fast amplitude oscillation appear almost simultaneously as those of the slow input oscillation, respectively. However, this situation does not always happen, and it highly depends on $\delta_\f$ in Eq.~\eqref{SL:main:input}. As pointed out in Ref. \cite{PM-MC-BL:87}, since $\sigma$ in Eq.~\eqref{SL:main} determines the convergence rate to the limit cycle, we then reason that the rate of $r_\f$ tracking $\sqrt{\sigma(t)}$ is also positively correlated with $\delta_\f$. Therefore, a larger $\delta_\f$ means a smaller phase lag. FIG.~\ref{Input}(c) demonstrates that a noticeable phase lag appears for a smaller $\delta_\f$ in comparison with FIG.~\ref{Input}(a).

After revealing the emergence of PAC, we investigate how this phenomenon depends on the model parameters. More precisely, we utilize the Modulation Index (MI) \cite{ABLT-RK-HE-NK:2010} to measure PAC intensity (see definition in the Supplementary Material). As we have reasoned that the fast amplitude $r_\f$ tends to track $\sqrt{\sigma(t)}$=$\sqrt{\delta_\f}\cdot\sqrt{1+{k_\RI}/{\delta_\f} \sin (\omega_\RI t)}$, it is clear that a larger ratio ${k_\RI}/{\delta_\f}$ implies larger fluctuations in $r_\f$ than a weaker one. That is, the PAC intensity increases with ${k_\RI}/{\delta_\f}$ (see FIG.~\ref{Input}(d)). Additionally, for a fixed a ratio ${k_\RI}/{\delta_\f}$, variations of $\delta_\f$ impact the fluctuations of $r_\f$: a larger $\delta_\f$ implies more intense PAC (see FIG.~\ref{Input}(d)). 

The input frequency $\omega_\RI$ also has a profound impact on the PAC intensity. A smaller $\omega_\RI$ means a more slowly varying $\sigma(t)$, which is easier for the fast amplitude $r_\f$ to track.  FIG.~\ref{Input}(e) shows that the PAC intensity decreases as $\omega_\RI$ increases. By contrast, the fast frequency $\omega_\f$ has little influence on the PAC intensity (see FIG.~\ref{Input}(f)). 

So far we have focused on the emergence of PAC in the architecture shown in FIG.~\ref{PAC}(b). That is, a slowly oscillating input to the fast population modulates the fast amplitude. However, some \textit{in vitro} studies have shown that besides high-frequency rhythms, low-frequency ones can also be generated locally in a single brain region \cite{WJA-BMI:2000,WMA-TRD:2003}. These heterogeneous rhythms then interact as in the architecture depicted in FIG.~\ref{PAC}(c). We now turn to the study of this architecture.

{\noindent \emph{PAC by local interactions: shaping of a brain region}} -- As depicted in FIG.~\ref{PAC}(c), a fast population may locally interact with a slow one. The two population dynamics~are 
\begin{align}
&\dot z_{\s}=(\delta_\s +\I v_\s-|z_\s|^2)z_\s+z_sf(z_\f),\label{dy:slow}\\
&\dot z_{\f}=(\delta_\f +\I v_\f-|z_\f|^2)z_\f+z_ff(z_\s),\label{dy:fast}
\end{align} 
where $z_\s=x_s+\I y_\s, z_\f=x_\f+\I y_\f$, and {$\delta_\s,\delta_\f>0$. The oscillators' natural frequencies are determined by $v_\s$ and $v_\f$, with $v_\f>v_\s$. The complex function $f(\cdot)$ describes how the two populations are interconnected. If the connection is unidirectional from the slow population to the fast one (a case studied in Ref. \cite{HA-FL-KC:2015}), i.e., $f(z_\f)=0$, then Eq.~\eqref{dy:fast} reduces to Eq. ~\eqref{SL:main:input} with input $u= f(z_\s)$.  In this case, our analysis above concludes that PAC emerges if $f(z_\s)$ is a function of the slow phase.

We next turn our attention to the more interesting case wherein the interaction between the two populations is bidirectional. In this Letter we simply consider $f(z)=kz$. Then, Eq. \eqref{dy:slow} and Eq. \eqref{dy:fast} become
\begin{align}
&\dot z_\s=\big((\delta_\s+kx_\f) +\I (v_\s+ky_\f)-|z_\s|^2\big)z_\s,\label{dy:slow:z}\\
&\dot z_\f=\big((\delta_\f+kx_\s) +\I (v_\f+ky_\s)-|z_\f|^2\big)z_\f.\label{dy:fast:z}
\end{align} In polar coordinates, $z_\s=r_\s e^{\I \theta_\s}$ and $z_\f=r_\f e^{\I \theta_\f}$. It can be observed that the terms $\sigma$ and $\omega$ (see Eq.~\eqref{SL:main}) that describe the amplitude and phase in the slow and fast population read as $\sigma_\s=\delta_\s+kx_\f$, $\omega_\s=v_\s+k y_\f$, and $\sigma_\f=\delta_\f+k x_\s$, $\omega_\f=v_\f+k y_\s$, respectively. Interestingly, the amplitudes and phases of the two populations become dependent upon one another due to the interconnection. Notice that such dependence is asymmetric due to the frequency difference.

\begin{figure}[t]
	\includegraphics[scale=0.7]{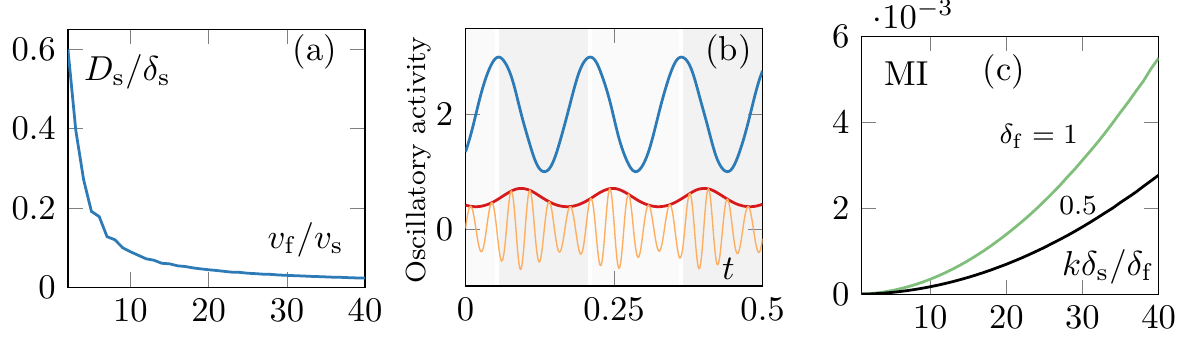}
	\caption{PAC by local interactions. (a) $D_s/\delta_\s$ with $D_s=\sup_t|| z_\s(t)-\hat z_\s(t)||$ measures the derivation of $z_\s(t)$ from the solution of its decoupled counterpart. As $v_\f/v_\s$ increases, the derivation shrinks dramatically ($k=5,\delta_\f=0.3$). (b) PAC emerges by local interaction between two populations as in FIG. \ref{PAC} (c), where the slow frequency is 6.5 Hz ($\theta$ band), and the fast one is 30Hz ($\gamma$ band). (c) PAC intensity (measured by the modulation index (MI)) increases with $k\delta_\s/\delta_\f$. Given $k\delta_\s/\delta_\f$, PAC becomes more intense for larger $\delta_\f$. }
	\label{PAC:two-nodes}
\end{figure}

We find that the slow oscillation remains relatively independent from the fast one. Since $v_\f>v_\s$, one can see that $z_\f$ oscillates faster than the $z_\s$. Applying averaging techniques developed in Ref.~\cite{SJA-VF-MJ:07} to the dynamics in Eq. \eqref{dy:slow:z}, we obtain the approximate guiding system $\dot {\hat{z}}_\s=(\delta_\s +\I v_\s-|\hat z_\s|^2)\hat z_\s$, which corresponds to the dynamics of the decoupled slow population. The solutions to Eq. \eqref{dy:slow:z} and the guiding system satisfy $\|z_\s(t)-\hat{z}_\s(t)\| \le  c {v_s}/{v_f}$ for some positive $c$. Loosely speaking, the slow trajectory ${z}_\s(t)$ fluctuates in the neighborhood of the trajectory of its decoupled counterpart $\hat{z}_\s(t)$. Such fluctuations decrease as the two frequencies separate (see Fig.~\ref{PAC:two-nodes}(a)). In contrast, Eq. \eqref{dy:fast:z} can be rewritten as $\dot z_\f=\big((\delta_\f+kr_\s \cos \theta_\s) +\I (v_\f+r_\s \sin \theta_\s)-|z_\f|^2\big)z_\f$. Similar to our earlier analysis, the fast amplitude $r_\f$ tends to track $\sqrt{\delta_\f+kr_\s \cos \theta_\s}$, which implies PAC.

Let the fast population in FIG.~\ref{PAC}(c) oscillate in the $\gamma$ band, and the slow population oscillate in the $\theta$ band (a choice compatible with empirical observations). FIG.~\ref{PAC:two-nodes}(b) shows that PAC emerges from the interaction of these two populations. The PAC intensity here also depends on the slow amplitude $r_\s$, which is determined by $\delta_\s$. Note that the connection strength $k$ also affects PAC intensity. Therefore, one can infer that the PAC becomes more intense as $k\delta_\s/\delta_\f$ increases (see FIG.~\ref{PAC:two-nodes}(c)).   

We point out that the architecture depicted in FIG.~\ref{PAC}(c) constitutes the basic organization of a brain region containing heterogeneous populations. This architecture can be easily extended to account for situations in which more than two frequencies coexist in a single brain region.

To corroborate the claim that our proposed architecture can reproduce PAC that resembles empirical signals in individual brain regions, we resort to \emph{in silico} experiments. As a common recording technique, local field potentials (LFP) are typically recorded by an electrode placed in the interested brain region to measure the local voltage variation that results from charge separation in the extracellular space. LFP signal analyses revealed that PAC takes place both in the striatum and the hippocampus \cite{TABL-MAT:08}. Following the method in Ref.~\cite{ABLT-RK-HE-NK:2010} (which we elaborate in the Supplementary Material), we generate synthetic LFP signals with diverse PAC intensity. We found that the model proposed in Eq.~\eqref{dy:slow}-\eqref{dy:fast} in the architecture depicted in FIG.~\ref{PAC}(c) can be tuned to be almost indistinguishable, after an initial transient, from synthetic LFP signals. We present an example in FIG.~\ref{synthetic}.
 
\begin{figure}[t]
	\includegraphics[scale=0.47]{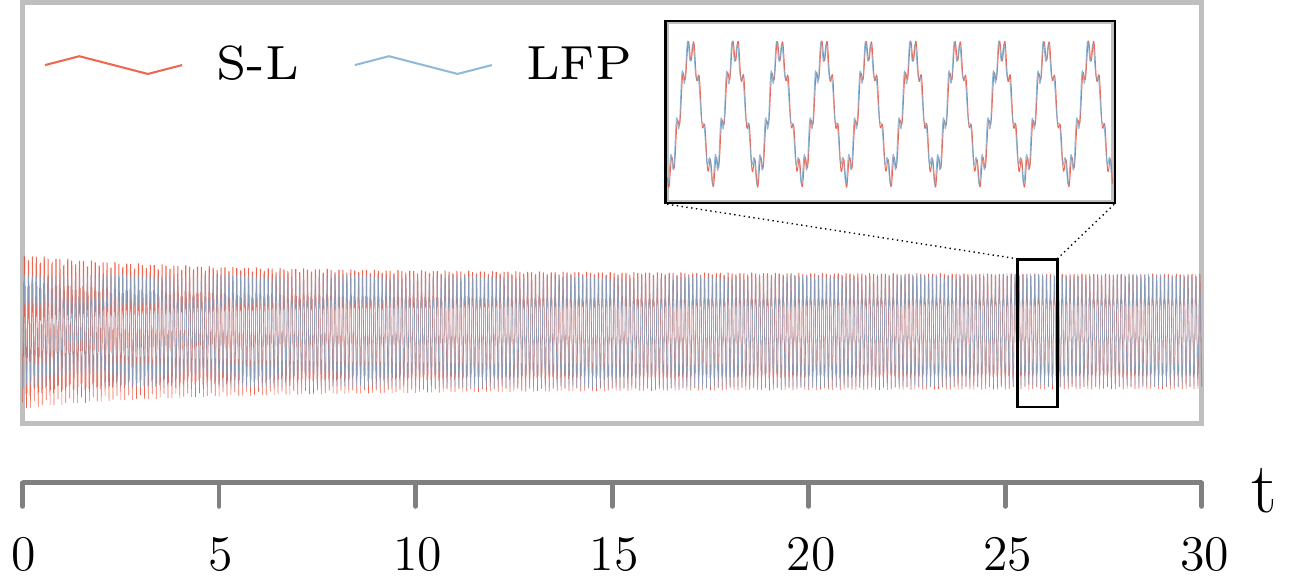}
	\caption{The SL model precisely reproduces synthetic LFP signals. Following previous empirical findings \cite{TABL-MAT:08}, in this simulation we have fixed the slow and fast frequencies to $10$ and $80$ Hz, respectively. We have also set $\delta_\text{s} = 0.217$ and $\delta_\text{f} = 0.038$ in the SL model (in blue), while the coupling $k$ is kept unitary. The integration of Eq. \eqref{dy:slow:z}-\eqref{dy:fast:z} was performed on Matlab with standard variable-step ODE45 solver. }
	\label{synthetic}
\end{figure}

Because numerous brain functions require synchronous activation of different brain regions, we now extend the single-region architecture discussed above (FIG.~\ref{PAC}(c)) to multiple coupled regions.

{\noindent \emph{Phase synchrony governs PAC across distant regions}} -- As a building block for more complex network structures, we consider a two-region clique as depicted in FIG.~\ref{PAC}(d), wherein interaction across brain regions is established by oscillations in the low-frequency range. Then, we show how cross-region phase synchrony contributes to integrate local high-frequency activities across long distances and to control the direction of information flow between regions.

Under the structure depicted in FIG. \ref{PAC} (d), the dynamics of the neural populations become
\begin{align}
&\dot z^p_{\s}=(\delta^p_\s +\I v^p_\s-|z^p_\s|^2)z_\s+k^p z^p_\s z^p_\f +k_c(z^{-p}_{\s}-z^p_{\s}),\label{cr-regi:dy:slow}\\
&\dot z^p_{\f}=(\delta^p_\f +\I v^p_\f-|z^p_\f|^2)z_\f+k^p z^p_\f z^p_\s,\label{cr-regi:dy:fast}
\end{align} 
where the superscript $p\in\{1,2\}$ is the region index ($-p=2$ if $p=1$ and $-p=1$ otherwise). Notice that, differently from the within-region connections across frequencies, the cross-region connection within the same frequency band is diffusive, which permits synchrony of slow oscillations. Here, $k^p$ and $k_c$ are the within- and cross-region connection strength, respectively. 

We find that the cross-region connection strength $k_c$ determines the phase synchrony between the two slow populations. One can reason that if the slow phases are synchronized, the fast amplitudes also behave coherently, provided there is local PAC in each region.  Let $\bar r^1_{\f}$ and $\bar r^2 _{\f}$ be the average amplitudes in the two regions. To measure the correlation of these two amplitudes, we calculate the phase locking value \cite{ABLT-RK-HE-NK:2010} defined as ${\rm PLV}=\frac{1}{T}\sum_{t=1}^{T}e^{\I(\Phi{\bar r^1_{\f}}(t)-\Phi{\bar r^2_{\f}}(t))},$ where $\Phi{r}$ is the phase angle of the amplitude $r$. Let $\Delta \omega_\s=|v^1_\s-v^2_\s|$ be the natural frequency of the two slow populations. The ratio  $k_c/\Delta\omega_\s$ determines the synchrony of the slow populations. FIG. \ref{PLV}(a) shows that the PLV increases with the ratio $k_c/\Delta\omega_\s$. Phase synchronization in the low-frequency range coordinates local high-frequency activities regulated by PAC, which is consistent with empirical studies such as Ref.~\cite{VNC-EG-SA:2014}. 

\begin{figure}[t]
	\includegraphics[scale=0.67]{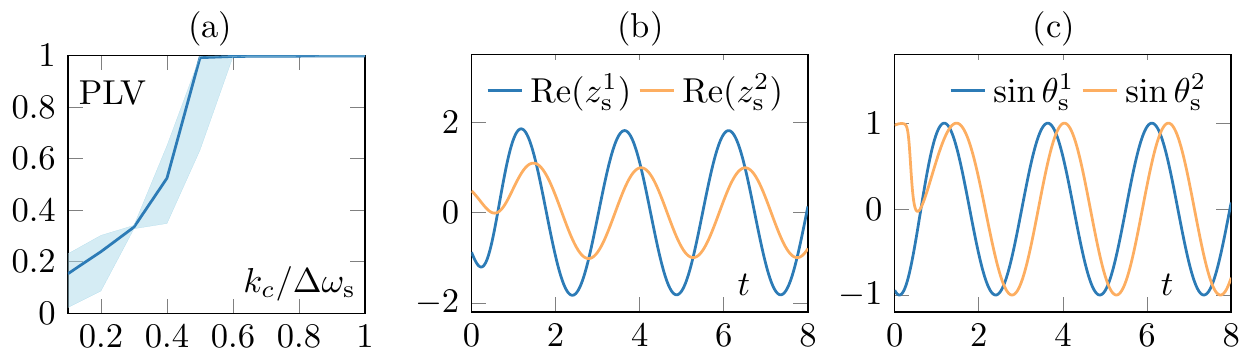}
	\caption{Roles of Low-frequency phase synchrony. (a) PLV of the fast amplitudes in the two regions against the ratio $K_c/\Delta \omega_\s$. The line indicated the mean of 1000 random realizations contained in the shaded area.  It can be inferred that the fast populations becomes more correlated as the synchrony of the slow oscillations increases. (b)-(c) Lead-lag relationship between the two signals: $z^1_\s$ precedes $z^2_\s$ if $v^1_\s>v^2_\s$ and vice versa.}
	\label{PLV}
\end{figure}

Finally, we put forth that our model may shed new light on the directionality of information flow across brain regions. Specifically, we show how natural frequency differences affect the lead-lag relationship of oscillatory rhythms between brain regions. It suffices to investigate the lead-lag direction in the low-frequency range, as we already know that slow oscillations govern communications across regions. Moreover, within each region the slow population remains relatively independent from the fast one. Thus it is sufficient to study the two diffusively coupled slow populations in FIG. \ref{PAC}(d), whose dynamics are approximately
\begin{align}
&\dot z^p_{\s}=(\delta^p_\s +\I v^p_\s-|z^p_\s|^2)z_\s+k_c(z^{-p}_{\s}-z^p_{\s}), \label{two_SL}
\end{align}
where $p=1,2$. In polar coordinates, $z^p_{\s}=r^p_{\s} e^{i\theta^p_{\s}}$, and Eq. \eqref{two_SL} can be rewritten as
\begin{align*}
&\dot r^p_{\s}=(\delta^p_\s-({r^p_{\s}})^2)r^p_{\s}+k_c(r^{-p}_{\s}\cos (\theta^{-p}_{\s}-\theta^p_{\s})-r^{-p}_{\s}),\\
&\dot \theta^p_{\s}=v^p_{\s}+k_c\frac{r^{-p}_{\s}}{r^{p}_{\s}}\sin(\theta^{-p}_{\s}-\theta^p_{\s}).
\end{align*}
For a strong connection $k_c$, the system converges to a solution with synchronized frequencies and constant but different amplitudes, i.e., $\dot \theta^1_{\s}-\dot \theta^2_{\s}=0$, $\dot r^1_{\s}=0$ and $\dot r^2_{\s} =0$. Letting $\theta^1_{\s}-\theta^2_{\s}=c_{12}$ yields
\begin{align*}
v^1_{\s}-v^2_{\s}-k_c\left(\frac{r^2_{\s}}{r^1_{\s}}+\frac{r^1_{\s}}{r^2_{\s}}\right)\sin c_{12}=0,
\end{align*}
which has a solution $c_{12}=\arcsin(\frac{v^1_{\s}-v^2_{\s}}{k_c \lambda})$ with $\lambda=\frac{r^2_{\s}}{r^1_{\s}}+\frac{r^1_{\s}}{r^2_{\s}}>0$ (the other solution $c_{12}=\pi-\arcsin(\frac{v^1_{\s}-v^2_{\s}}{k_c \lambda})$ is unstable). The system converges to a solution wherein the phase $\theta^2_{\s}$ lags behind (resp. leads) $\theta^1_{\s}$  if $v^1_{\s}>v^2_{\s}$ (resp. $v^1_{\s}<v^2_{\s}$), which we illustrate in FIG.~\ref{PLV}(b)-(c). Suppose $v^1_{\s}>v^2_{\s}$, then we have that $z^2_\s(t)=r^2_\s\cos(\theta^2_\s(t))+\I \sin(\theta^2_\s(t))=\frac{r^2_\s}{r^1_\s}z^1_\s(t-\frac{c_{12}}{\omega})$ with $\omega$ being the synchronized frequency, which means that $z^2_\s$ lags behind $z^1_\s$ by $\tau={c_{12}}/{\omega}$.  This behavior may suggest that the former acts as the information sender, and the latter acts as the receiver, since the lead-lag relationship typically encodes the directionality of the information flow \cite{NG-ZA-NVV:2008,MJY-LU:2015}.

To summarize, we postulate that synchronization supports effective communication, and that the natural frequency differences determine the information flow directionality. We conjecture that the brain is capable of controlling this directionality effectively by manipulating the natural frequencies of neural populations. Further experimental studies are needed to investigate these conjectures. We remark that our findings provide some useful insights regarding the inference of directionality in information flow. As low-frequency oscillations are much more engaged in cross-region communication than the high-frequency counterparts, the latter may be filtered out from recorded signals when analyzing information flow between regions. 

In this Letter, we have focused on modeling cross-frequency phase-amplitude coupling (PAC). A cross-region structure for interaction of multiple brain regions has also been proposed. The combination of this interconnection structure and our model may pave a way to study other cross-frequency coupling apart from PAC, same-frequency phase synchrony, and their interplay.

\begin{acknowledgments}
	This work was supported in part by awards ARO W911NF-18-1-0213, ARO 71603NSYIP, and NSF NCS-FO-1926829.
\end{acknowledgments}

\bibliographystyle{apsrev4-2}
\bibliography{\alias,\FP,\Main,\New}

\section{Supplemental Material}
\textbf{Modulation Index (MI).} To measure the intensity of phase-amplitude coupling (PAC), we employ the Modulation Index (MI) \cite{ABLT-RK-HE-NK:2010}. In the following, we summarize the steps to compute the MI.

We first derive the phase-amplitude distribution $P$ of two complex signals -- a slow one $z_\s(t)=x_\s(t)+\I y_\s(t)$, and fast one $z_\f(t)=x_\f(t)+\I y_\f(t)$ -- as follows:
\begin{enumerate}
	\item[(1)] We extract the time series of the phases of $z_\s(t)$. For each time instant $t$, the phase is $\theta_\s(t)$ in $z_\s(t)=r_\s(t) e^{\I \theta_\s(t)}$.
	\item[(2)] Extract the time series of the amplitude envelope of $z_\f(t)$. Analogously, the amplitude at each time instant $t$ is $r_\f(t)$ in  $z_\f(t)=r_\f(t) e^{\I \theta_\f(t)}$.
	\item[(3)] Divide the phases $\theta_\s(t)$ into $n$ bins, where $n$ can be any positive number (we use 18 in the our text). For each bin $i$, calculate the mean of $r_\f$, denoted by $\langle r_\f \rangle_i$.
	\item[(4)] Normalize the mean amplitude in each bin by
	\begin{align*}
		P(j)=\frac{\langle r_\f \rangle_i}{\sum_j^n \langle r_\f \rangle_j}.
	\end{align*}
\end{enumerate}
Notice that the phase-amplitude distribution satisfies $P(i)\ge 0$ for any $i$ and $\sum_{j=1}^nP(j)=1$, akin to a probability mass function. The intensity of PAC can be inferred intuitively by visual inspection of the plot of $P$ over all the bins. Nevertheless, to provide a quantitative measure of PAC intensity, the Modulation Index is defined as 
\begin{align*}
	{\rm MI} =\frac{D_{KL}(P,U)}{\log (n)}
\end{align*}
where $D_{KL}(P,U)=\log (n)-H(P)$, $H(P)=-\sum_{j=1}^{n} P(j)\log (P(j))$, and $U$ represent the uniform distribution. Loosely speaking, the Modulation Index is a scaled Kullback-Leibler distance of the phase-amplitude distribution from the uniform distribution, satisfying $0\le \text{MI} \le 1$. $\text{MI}=1$ means that $P$ is a Dirac-like distribution, while $\text{MI}=0$ is equivalent to a uniform distribution. Thus, for any signal, the larger the MI is, the more intense the PAC.  

\textbf{Generating synthetic local field potential (LFP).} Following \cite{ABLT-RK-HE-NK:2010}, we generate synthetic LFP signals as

\begin{align*}
	x(t)=r_\f(t) \sin (2\pi \omega_\f t)+r_\s \sin(2\pi \omega_\s t)+W(t),
\end{align*}
where $r_\f(t)$ is the amplitude envelope of the fast oscillation, $r_\s$ is a constant representing the amplitude of the slow oscillation, and $W(t)$ is a Gaussian white noise.  To ensure that the synthetic signal contains phase-amplitude coupling, $r_\f(t)$ is defined as
\begin{align*}
	r_\f(t)= \bar r_\f \frac{\mu \sin(2 \pi \omega_\s t)+2-\mu}{2},
\end{align*}
where $\bar r_\f(t)$ is a constant determining the maximum amplitude of the fast oscillation, and $\mu$ is the fraction of the amplitude envelope modulated by the slow oscillation. The PAC intensity is thus controlled by the parameter $\mu$. LFP signals with diverse PAC intensity can be generated by manipulating $\mu$.

\end{document}